# Formal Hierarchies and Informal Networks: How Organizational Structure Shapes Information Search in Local Government

Travis A. Whetsell
Department of Public Policy and Administration
Steven J. Green School of International and Public Affairs
Florida International University
11200 SW 8th Street, PCA 260A
Miami, FL 33199
travis.whetsell@fiu.edu

Alexander Kroll
Department of Public Policy and Administration
Steven J. Green School of International and Public Affairs
Florida International University
11200 SW 8th Street, PCA 351B
Miami, FL 33199
akroll@fiu.edu

Leisha DeHart-Davis
School of Government
The University of North Carolina at Chapel Hill
400 South Road
Chapel Hill, NC 27514
ldehart@sog.unc.edu

**Abstract:** Attention to informal communication networks within public organizations has grown in recent decades. While research has documented the role of individual cognition and social structure in understanding information search in organizations, this article emphasizes the importance of formal hierarchy. We argue that the structural attributes of bureaucracies are too important to be neglected when modeling knowledge flows in public organizations. Empirically, we examine interpersonal information seeking patterns among 143 employees in a small city government, using exponential random graph modeling (ERGM). The results suggest that formal structure strongly shapes information search patterns while accounting for social network variables and individual-level perceptions. We find that formal status, permission pathways, and departmental membership all affect employees' information search. Understanding the effects of organizational structure on information search networks will offer opportunities to improve information flows in public organizations via design choices.

# Introduction

Research in public administration has increasingly drawn attention to the salience of interpersonal communication networks within public organizations (Moynihan and Pandey 2008; Siciliano, 2015; Nisar and Maroulis, 2017). To accomplish daily work tasks, individuals often lack the requisite information necessary to perform effectively. They seek out information from other individuals whom they perceive to have access to important knowledge. With the increasing informational intensity of the workplace, public organizations depend on the development of communication networks of individuals that span teams, departments, and formal lines of authority.

Interpersonal networks often emerge from informal collaboration and lateral coordination among individuals within organizations (Berry et al. 2004; Isett et al. 2011). Research on networks within organizations has examined the importance of social processes, such as reciprocity and transitivity, as well as the effect of individuals' position in social networks (Krackhardt and Hanson 1993; Contractor and Leonardi 2018). Being at the periphery of a network is associated with negative attitudes toward one's work (Porter et al. 2019), whereas being in a position that bridges organizational sub-groups is correlated with positive outcomes (Maroulis 2017). With regard to knowledge networks, research similarly emphasizes the importance of informal network structure. Better connected individuals are able to leverage more organizational knowledge (Oparaocha 2016; Paruchuri and Awate 2017; Tasselli 2015).

In this article, we take different perspective on knowledge networks. In line with previous work, we employ a network-based conceptualization of communication: that is, forming ties for the purpose of information search. Unlike most other work we do not focus only on drivers of tie formation based on network structure or attributes of individuals. In fact, we argue quite the opposite: The *formal* structure of the organization and its hierarchy shape *informal* communication networks in important ways that have largely been neglected in the literature. In line with previous work (Krackhardt and Hanson 1993; McEvily, Soda, and Tortoriello 2014), we consider networked, social structures to be informal and designed or engineered organizational structures and processes as formal. We recognize that studying informal networks in organizations itself is partially a response to the historical dependence on formal structures (Granovetter 1985; Krackhardt and Hanson 1993), and we do not suggest shifting back from the informal to the formal. Rather, we propose that we can improve our understanding of intra-organizational networks by accounting for the individuals' place within a formal structure, thereby, connecting the formal with the informal.

While the "missing link" between formal organization and informal social structures has been occasionally discussed in the organizational sciences (McEvily, Soda, and Tortoriello 2014; Hunter, Bentzen, and Taug 2020), we argue that organizational structures are particularly influential in public administration due to its reliance on rules, the hierarchy, and formal processes (Hill and Lynn 2015). In essence, different configurations of the bureaucracy may affect public managers' work and its outcomes, even in highly networked environments (O'Toole and Meier 2010). This is likely to be even more true for generalist jurisdictions, where distinct departments take on very different functions, compared to specialist agencies, where tasks are narrower and sub-units more homogenous. Interestingly, however, most network research either focuses on private firms or, in the public sector, the highly specialized school system. We extend this work by studying more generalist local governments, which constitute one of the most common cases within public administration but have received little attention in the social network literature.

To investigate the connection between formal organization and informal social structure, we develop three hypotheses about the role of formal status, permission pathways, and departmental membership in explaining informal information search among government employees. Empirically, we show that all three formal-structural factors are important determinants of information search, controlling for the important attitudinal constructs of trust, commitment, and self-efficacy, as well as important social network processes such as reciprocity, transitivity, and popularity.

This article models these effects using exponential random graph modeling on a sample of 143 employees (reflecting a response rate of 92%) across multiple departments in a small city government. We estimate the effects on information seeking behavior while controlling for a set of social-structural network variables as well as important cognitive variables. The results suggest that structure has important shaping effects on information search. Tie formation in the information search network is both constrained and driven by several aspects of organizational structure. The results suggest that formal structure has important implications for scholars and practitioners concerned with understanding the antecedents of information search in public organizations, suggesting that organizational design and interventions associated with formal structure may stimulate the emergence and maintenance of robust communication networks necessary to perform work tasks in the local government context.

## Literature Review

In this section, we first provide a brief overview of the literature on *intra*-organizational networks in public administration. While certainly growing, this body of research is still underdeveloped, especially compared to the work on *inter*-organizational networks. Next, we review the literature on information searching and sharing in public organizations to synthesize the state of research and identify our research gap. Finally, we develop hypotheses about the importance of organizational structure for shaping information search in social networks.

**Intra-Organizational Networks in Public Administration**

The study of social networks within and between public organizations is an emerging subject in public administration research. During the latter half of the 20[th] century, scholars developed methods for analyzing social relations between individuals in terms of graph theory and networks. The term "network" first appeared in the titles, abstracts, and keywords of articles in *Public Administration Review* in the mid-1980s. However, the substantive application of network analysis did not begin to occur until much later, with agenda setting work (e.g. O'Toole 1997) appearing in the mid to late 1990s (Hu, Khosa, Kapucu 2016).[1] Since then, a great deal has been uncovered about the resolution of public problems through social and organizational networks (Provan and Milward 2001; Berry et al. 2004; Provan and Kenis 2008; Isett et al 2011). For example, the concepts of policy networks, collaborative networks, and network governance emerged to provide theoretical explanation for the increasingly complex patterns of interaction between numerous actors around public programs, policies, or problems (Kapucu, Hu, and Khosa 2017).

Much of the research on networks in public administration has occurred at the *inter*-organizational level, with fewer articles exploring *intra*-organizational social networks between individuals within public organizations. Kapucu, Hu, and Khosa (2017) suggest that merely 14% of public administration network studies have used the individual as the level of analysis.

---

[1] A Web of Science search ((SU="public administration") AND (TS="network analysis")) show the term network analysis did not appear in the searchable text of a general public administration journal until 1997.

Similarly, Isett et al. (2011) state that informal interpersonal networks are an understudied area. Information seeking networks of individuals are categorized as informal networks, as opposed to formal hierarchical relations or formal inter-organizational relations that involve some contractual exchange of resources.[2]

Intra-organizational networks are often defined in terms of the relationships between individuals within organizations. Such networks emerge as a result of complex communication patterns between individuals, where the exchange of information or resources is the generic criterion for quantification of a relationship (tie) between two individuals in a network (Monge and Contractor 2003). As the number of ties increases between individuals, and as secondary ties begin to form between their partners, a network structure begins to emerge. Network structure can vary in terms of size, density, diameter, and centralization, etc. with consequences for the flow of information and resources in a network (Wasserman and Faust 1994). As networks increase in size and density, they often begin to form a core-periphery structure, where the distribution of ties is concentrated around a few very well-connected individuals or groups (Borgatti, Everett, and Johnson 2015). A ubiquitous phenomenon known as preferential attachment characterizes highly skewed distributions of network ties concentrated among few very well-connected nodes, observed in numerous types of networks across physical, biological, and social networks (Barabasi and Alber 1999; Newman 2001).

An individual's position within the network has important influences on access to resources and perceptions of social status within the network. For example, individuals who occupy central positions in the network often benefit from enhanced access to information, resources, and the popularity and prestige that derive from the social capital associated with such access (Lin 1999). Further, as routines of exchange develop among network actors, mechanisms of social interaction emerge to safeguard and maintain the structure and functioning of the network (Jones, Hesterly, and Borgatti 1997). Thus, Coleman (1994) posits a reciprocal process between structure and actor, where social structure has a downward influence on individual actor-level behavior, but actor-level behavior has upward influence back on structure, i.e. Coleman's Boat.

---

[2] A topic search for intraorganizational or interpersonal network within the subject category of public administration yielded only 24 results in the Web of Science data base (accessed 01/23/2020).

While the literature on intra-organizational and interpersonal networks within public organizations is limited, there are several important studies that examine the effects of such networks on variables of interest to management and organizational behavior scholars, including turnover intention (Moynihan and Pandey 2008), organizational commitment (Siciliano and Thompson 2018), resource sharing (Fusi, Welch, and Siciliano 2019), innovation (Nisar and Maroulis 2017; Zandberg and Morales, 2019), and individual performance outcomes (Siciliano 2017).

Among others, this research has shown that organization-internal networks act as "sticky webs" that keep people in the organization, while external networks act more like "trampolines" to the next organization (Moynihan and Pandey 2008). Further, it has documented the social dependencies between individuals that shape perceptions and attitudes as well as differential effects in advice versus friendship networks (Siciliano and Thompson 2018). In a study of performance outcomes of schools, including network measures significantly improved the variance in scores explained by the models for reading as well as mathematics scores (Siciliano 2017).

**Information Searching and Sharing in Public Organizations**

Studies on information seeking and sharing in public organizations highlight the importance of interpersonal networks in facilitating the flow of knowledge for the accomplishment of work tasks. Binz-Scharf, Lazer, and Mergel (2012) apply the knowledge-based view (KBV) of organizations (Grant 1996) to analyze resource exchanges in an interpersonal network of forensic laboratory workers. The KBV treats knowledge as a resource critical to the performance of organizations. As Nonaka (1994:15) suggests, information and knowledge can be distinguished in the following manner: "information is a flow of messages, while knowledge is created and organized by the very flow of information, anchored on the commitment and beliefs of its holder." In this sense, information provides the material basis for the construction of theory with the pragmatic aim of guiding some action. However, because knowledge is often tacit rather than explicitly codified in organizational files, employees expend considerable effort on information search activities (Nonaka 1994; Polanyi 1996). Interpersonal communication networks facilitate the sharing of both tacit and explicit knowledge for the completion of work tasks. As Binz-Scharf, Lazer, and Mergel (2012) show, interpersonal networks are critical to the functioning and performance of knowledge-intensive public organizations. This research suggests that elements of the knowledge-based view of the firm may be generalized to public organizations.

While several studies have examined the effects of networks on outcomes of interest in the workplace, fewer studies have examined the antecedents of intraorganizational and intrapersonal network formation. Nisar and Maroulis (2017) studied information seeking in interpersonal communication networks of teachers in a public high school. Their results suggest that street-level bureaucrats tend to seek out information from individuals who use their discretion to experiment with new innovative practices in the workplace. Siciliano (2016) examines advice networks of teachers in five schools. His results suggest that the expertise of individuals becomes an important factor in being sought out for advice, specifically when the domain of activity is knowledge explicit. Conversely, he found the opposite to be true in less knowledge intensive domains of activity.

While the drivers of information sharing in public organizations are manifold (Yang and Maxwell 2011), we notice a predominance of individual-level factors, particularly cognitive variables, in social network studies. One such factor is trust. Mayer and colleagues provide the seminal definition of trust in organizations, as "the willingness of a party to be vulnerable to the actions of another party based on the expectation that the other will perform a particular action important to the trustor, irrespective of the ability to monitor or control that other" (1995, p. 712). Trust figures prominently into research on information exchange at the individual (Levin and Cross 2004), intra-organizational (van de Bunt, et al, 2005), and inter-organizational levels (Tsai and Ghoshal 1998; Shazi, Gillespie and Steen 2015). The most common explanation for trust and information exchange focuses on trust in information sources, where sharing is facilitated if the information source is perceived as reliable (Levin and Cross 2004; Shazi, Gillespie and Steen 2015). Trust can also foster information sharing due to reducing transaction costs between sender and receiver (Dawes, Cresswell, and Pardo 2009). As Yang and Maxwell (2011) suggest, when information is viewed as an asset within a broader organizational power game, withholding it from others is often a rational strategy for competitive advantage. Similarly, Dawes, Cresswell, and Pardo (2009), citing Jones, Hesterly, and Borgatti (1997), suggest that transactions costs associated with interpersonal communication increase when trust is low. Thus, trust substitutes for more costly organizational structures designed to prevent exploitation by increasing monitoring and oversight controls.

Another factor is self-efficacy, which is an individual's belief in their own capabilities given a specific domain of interest, such as in the performance of work tasks (Lunenburg 2011). Self-efficacy has been examined as an antecedent for various outcomes of interest in several articles in public administration most of which have been published in the last ten years (George et al. 2018; Jacobsen and Andersen 2017; Wright 2004). Self-efficacy effects on information search may materialize via two opposing logics: Individuals high in self-efficacy are less concerned with appearing incompetent to others due to confidence in their own abilities, suggesting that individuals higher in self-efficacy may be more willing to seek out knowledge from their coworkers. Conversely, individuals lower in self-efficacy may hesitate to reach for the costs of doing so, such as reputation damage and loss of self-esteem (Johnson, et al, 1995). Alternatively, it may be that individuals who have higher self-efficacy may feel themselves less likely to require advice or information from others. Siciliano (2015, 2016, 2017) examined the effects of self-efficacy on knowledge seeking activities in a set of network studies, and he overall documents mixed and null findings. However, Siciliano's results may be contingent on the context of the public-school workplace, where teachers have considerable autonomy and independence in the accomplishments of day-to-day teaching responsibilities. Analyzing self-efficacy in settings where employees require more interaction to accomplish work tasks may yield different results.

A third cognitive factor that is known to shape behaviors such as information sharing and search is employees' organizational commitment. This term is widely understood as "the affective attachment to the organization, perceived costs associated with leaving the organization, and obligation to remain with the organization" (Meyer and Allen 1991, 64). We suspect that commitment will foster social exchange around information. Both behaviors – searching for information and providing it when others reach out – will require additional effort from the employee, which can be more likely expected of committed workers. In many instances, information search constitutes a type of extra-role behavior, where employees need to look outside of the formal hierarchy to obtain information required to solve nonroutine problems. Similarly, helping others when they search for information is often comparable to organizational citizenship behavior, which employees committed to the organization's mission will engage in. Siciliano (2017) shows that commitment is a consequential variable in education networks. Research also suggests that public employees' identification with their organization fosters the effectiveness of knowledge sharing but had no impact on the knowledge-sharing intensity (Willem and Buelens 2007). Overall, we believe that organizational commitment will reinforce information seeking behaviors.

**Organizational Structure and Information Search**

A general ontological principle of social networks is that they constitute complex phenomena that emerge from local processes of self-organization between individuals, rather than emerging purely from formal organizational structure (Comfort 1994; Miller and Page 2009). However, organizational structure remains an important element in the development and shape of social networks (Cross, Borgatti and Parker 2002; Agranoff 2006; Eglene, Dawes, and Schneider 2007). Organizational structure in this respect pertains to the social architecture that arranges individuals and groups and delineates relationships between them (Tolbert and Hall 2009, 20; Hall 1999, 47). Structure is formal in that it is explicitly developed and sanctioned by the organization (Pugh, Hickson and Vinings 1968).

We focus on the location of individual members within formal structure in altering information search in intra-organizational networks, which (along with other forms of organizational structure generally) is an understudied topic in social networks (Johnson et al. 1995; Hunter et al. 2020). Three structural attributes of the individual are examined: formal status, permission pathways, and departmental membership.

*Formal Status*

Given the organizationally dependent nature of information search within public organizations, it is reasonable to question how formal status influences information search.[3] As Krackhardt (1990) suggests, those in higher positions of formal authority interact with more individuals and deal with more issues, often acting as bridges between disconnected others in a network. As he proposes, formal power influences *informal* network power, the cognitive accuracy with which others perceive the network, and the reputational power of others in the network. Individuals located upwards in the hierarchy are likely to have access to information relevant to the completion of work tasks both within and between departments (Kahn and Kram 1994). They also are likely to have the "last word" (decision-making authority), so that employees reaching out can be more certain the information they receive is final or has been vetted and approved (Allen and Cohen 1969; Wager 1972; Galbraith 1974).

---

[3] In related research, experimental evidence related to hierarchy and networks has examined the location of a central person in a network and its influence on communications patterns. Bavelas (1950) and Leavitt (1951) found that networks with communications funneled through centralized individuals were higher performing. Guetzkow and Simon found that centralized network structures performed more efficiently, but only if self-organized and not pre-ordained (1955). Mulder examined centralized decisionmaking and observed that centralized decision structures were higher performing (1960). And a comparison of flat versus tall decision structures found tall decision structures to be higher performing than flat ones, with no significant difference in tall and flat structures on decision speed (Carzo and Yanouzas 1969). However, not all research supports these linkages. In particular, some scholars have found that centralized network communications structures such as those found in hierarchies exert mixed (Shaw 1964) or no influence (Shore, et al, 2015; Maroulis et al 2020) on performance outcomes.

The dynamics of information search differ for individuals lower in the hierarchy. Generally speaking, hierarchical structure creates acceptable pathways for information flow in which lower status individuals are supposed to reach upwards to their immediate supervisor in seeking information (Bavelas 1950; Daft, Murphy and Willmott 2010). To circumvent the hierarchy and communicate over one's supervisor, or to reach out to peers, disrupts the status quo (Hage 1965), particularly if it involves acting outside one's functional boundaries (Burns and Stalker 1961). To occupy a hierarchically lower status position also reduces psychological safety by creating fears of retribution (Kish-Gephart et al., 2009), which may reduce information search and sharing in the organization. In support of these propositions, an experiment with hierarchical structures produced fewer communications upward and laterally than more democratic structures (Lyle 1961). This does not mean that employees, such as street-level bureaucrats, would not communicate a great deal with similar others. But our point is that relative to lower-level positions, higher-status individuals – by organizational design – need to reach out within and across units to manage effectively, and they serve as attractive sources of important information for others.

> *Hypothesis 1. Formal status is positively associated with the formation of information search ties.*

*Permission Pathways*

According to formal-organizational theory, "interactions are by design" (McEvily et al 2014, 306). That is, the way in which individuals work together follows formally defined processes and communication rules. While the role of rules and in shaping organizational behavior have been prominently featured in public administration research (e.g., DeHart-Davis 2017), formal rules that govern interactions among work groups and the movement of information are also at the forefront of process management scholarship (Nadler et al. 1997). Interestingly, formal processes are a factor often turned to in order to induce organizational reform, especially when reengineering strategies involve information technology (Davenport 1993). While network research has shown that social interactions can deviate from organizational design, our point is that formalized lines of interaction and authority may still be a useful – yet widely undervalued – predictor of informal information exchange.

Social network research has suggested that networks often exhibit *multiplexity*, where, for example, networks based on advice ties are often dependent on other sorts of underlying relations such as friendship ties (Lazega and Pattison 1999). As Siciliano suggests (2015:551) "Multiplexity is an important concept in the literature on intraorganizational networks given the tendency for formal roles (e.g., status, position) to overlap with informal roles." Extending this logic to the present case, we hypothesize that information search relationships may depend to some extent on existing organizational authority, such that individuals will tend to seek out information from those with whom they also have existing permission-based relations. If an individual frequently has contact with another in order to gain permission for a given work task, then they may also be more likely to seek that person out for information more generally. Further, such individuals may hold positions of authority precisely because they have information relevant to the effective completion of particular tasks. Thus, we advance the following hypothesis.

> *Hypothesis 2. Permission network ties are positively associated with the formation of information search ties.*

*Departmental Membership*

While the research on information searching and sharing has emphasized the social, informal dynamics behind such behaviors, we argue that organizational structure creates spaces in which these interactions occur. Departments, one type of organizational structuring, have been shown to be important prisms through which employees view the entire organization (Kroll, DeHart-Davis, Vogel 2019). Employees within the same department function within the same operating environment, and thus are more likely to have shared understandings of workplace norms, values and expectations (Ginsburg et al 2009; Schaubrook, et al, 2012). Furthermore, organizations are typically structured in ways that place task-interdependent individuals in work units (Thompson 1967; Galbraith 1974), making it highly likely that they turn to each other when they need information rather than look outside. Since departments impose order on lateral communication, we infer more specifically that departmental membership should also play a critical role in guiding information search (Cross, Rice and Parker 2001). In support of this argument, Johnson and colleagues (1995) found that individuals with an interpersonal dependence on others, which is a logical feature of individuals within a department, were more likely to seek information from others. As Kleinbaum et al. put it, "structure itself induces a great deal of interaction," and accordingly found higher levels of communication among dyads within the same business units (2013).

Using insights from the social network literature, intra-departmental relationships can be conceptualized as a type of actor-attribute based homophily, where a node is more likely to form ties with nodes that share the same attribute (McPherson, Smith-Lovin, & Cook, 2001). However, departmental homophily is distinct from the self-organizing "birds of a feather flock together" type and is instead induced from the top-down structure of the organization. Further, information search is more functional in nature, based on the resolution of a workplace problem rather than being purely derived from organic social processes. Different types of departmental structures may be more likely to produce within versus between departmental ties. Finally, variability of information search patterns shaped by departmental structure may be even greater in the generalist local government setting, where departments within the city are often vastly different in size, levels of hierarchy, and culture. Accordingly, we expect that:

> *Hypothesis 3 Departmental membership is positively associated with the formation of information search ties.*

## Research Design

### Data Collection

To test the relationship between formal structure, social dynamics and tie formation/information search, we emailed a Qualtrics survey link to all 155 employees of a small coastal local government in a Southeastern state. The survey was administered in October 2019 and remained open for two weeks. The response rate was 92 percent (n=143/N=155).[4] Two individuals only partially completed the attitudes portion of the survey, which is why their scores were imputed based on the median values of the sample.[5] The survey sample represents the city workforce in age, gender, and departmental representation.

The survey instrument used the roster method to generate social network data (Wasserman and Faust 1994; Wald 2014; Perry, Pescosolido, Borgatti 2018). The small city government context is useful for addressing the network boundary specification problem by restricting potential network actors to the common characteristic of employment within the organization (Nowell et al. 2018). Thus, the survey asked about the occurrence and nature of interactions between survey participants and the full roster of every other employee in the city.[6]

**Variables**

To identify network interactions between organizational members, survey participants were asked to indicate whether they sought out each organizational member for "information from this person to do my job." The survey item about seeking information established interactions between employees based on acquiring knowledge about specific day-to-day work tasks. While we kept the language broad, we distinguished it in the survey from discussion seeking and permission seeking. The language "seek out" was used to establish directionality to the network ties. For ease of survey readability, organizational members were divided into sections by department. This survey item provided the basis for constructing the information search network, where information ties within the network serves as our dependent variable. Exact wording for all survey items can be found in Online Appendix 1.

---

[4] Several factors contribute to the high response rate. First, as an incentive to participate, the city offered the chance for one survey participant to win eight hours of vacation. The winner was randomly selected by the research team. The winner was given the option of not having their name announced to the broader organization. A second factor is the reputation of the administering university, which is known, trusted and respected by local government employees.

[5] We chose to impute these observations because they completed the network portion of the survey, and ERGM analysis does not permit missing observations in the independent variables. While median imputation may bias standard errors on the attitude controls, there was very little difference on ERGMS with/without these two nodes.

[6] Names were displayed in alphabetical order. To test the possibility that the ordering of names might privilege individuals with last names earlier in the alphabet, we ran a correlation between indegree centrality (measured as the total number of times an individual is indicated as being sought out) and name order (numbered 1 through 155). The correlation was 0.06, indicating that roster ordering was not a factor in the selection of information seeking ties. We also ordered the names by department, finding no general pattern that would indicate a bias toward selecting individuals appearing earlier in the roster.

Organizational trust was measured using three items related to supervisor, team, and organization specific trust, which were adapted from Kroll, DeHart-Davis, and Vogel (2019). The scale responses ranged from 0 (Strongly Disagree) to 6 (Strongly Agree). Cronbach's alpha for the items is 0.92. Organizational commitment was measured using three items, adapted from Meyer and Allen 1991, including turnover intention (a derivation of continuance commitment), guilt about leaving the organization (normative commitment), and happiness regarding the prospect of remaining with the organization (affective commitment). These three items were selected for the survey to tap concepts across Meyer and Allen's three dimensions while also being parsimonious in the number of survey items presented to research participants. The Cronbach's alpha for these items is 0.85. Self-efficacy (the perception that goals are achievable despite difficulties) was measured using three items from Chen, Gully and Eden (2001). The scale responses ranged from 0 (Strongly Disagree) to 6 (Strongly Agree). Cronbach's alpha for the items is 0.92. Each set of variables was aggregated into principal components for use in the models; all three principal components produced eigenvalues above one, and the analysis produced a single component for each variable set. When factor analyzing all nine items together, they load on their original three factors with no significant cross-loadings (factor table is included in Online Appendix 2).

Our organizational-structural variables were constructed based on administrative data to avoid issues related to common-source bias. To operationalize Hypothesis 1, regarding formal status, we employ two variables to tap into the concept of formal status. Administrative data allowed us to calculate the "hierarchical status" for individuals, defined as the number of steps to the top position (in this case, the city manager). The variable was then reverse coded to account for higher rather than lower status. The administrative data also permitted identifying whether individuals had "supervisory status," which served as our second measure of formal status. To operationalize Hypothesis 2, regarding permission pathways, we generated a separate permission network alongside the information network, which asked participants to identify the individuals they seek out when they need "permission" to complete work tasks. The analysis includes the effect of tie formation in the permission network on tie formation in the information network by including an edge covariate in the analysis (edgecov). To operationalize Hypothesis 3, regarding department effects, we identified departments based on the city budget document for FY2019-2020 and used department membership as a categorical node attribute to control for department specific effects, both for receiver and sender effects (receiver.factor/sender.factor). We also include a nodematch term for department to identify assortative mixing, e.g., homophily, between individuals in the same department.

We also include in Table 1 four descriptive network statistics that are often useful in describing individual level network position. While these are not central to the hypothesis testing, they are important to an initial descriptive analysis of networks. In-degree is the number of incoming ties a node receives; out-degree is the number of ties a node sends; betweenness centrality shows the number of times a node is on a path between other nodes (normalized); transitivity is the clustering coefficient of the nodes, which represents the degree to which the node is embedded within a cluster of nodes (using transitive triads as the measure) (Wasserman & Faust 1994).

**Table 1. Descriptive Statistics**

| Variable | n | mean | sd | min | max |
|----------|-----|------|------|-------|-------|
| Self-efficacy | 143 | 0.00 | 1.00 | -5.18 | 0.73 |
| Trust | 143 | 0.00 | 1.00 | -3.04 | 1.27 |
| Commitment | 143 | 0.00 | 1.00 | -2.69 | 1.16 |
| Hierarchical status | 143 | 3.34 | 1.45 | 1.00 | 7.00 |
| Supervisory status | 143 | 0.32 | 0.47 | 0.00 | 1.00 |
| Indegree | 143 | 12.29 | 10.18 | 0.00 | 48.00 |
| Outdegree | 143 | 12.29 | 12.27 | 0.00 | 67.00 |
| Betweeness | 143 | 0.44 | 0.23 | 0.00 | 1.00 |
| Transitivity | 143 | 0.01 | 0.02 | 0.00 | 0.17 |

The ERG model allows for numerous model terms per variable of interest. We decided to include the same model terms for each variable: a term for direct tie sender effect, direct tie receiver effect, and a term which captures the effect of assortative mixing, such as homophily or heterophily: nodematch for factor variables or absdiff/diff for continuous variables.

Finally, we include structural network terms to account for the density of the network (edges); reciprocity in tie formation (mutual); transitivity, measured through directed geometrically weighted edgewise shared partner distribution (DGWESP); popularity, measured through geometrically weighted in-degree distribution (GWIDegree); and activity, measured with two out-degree distribution terms (ODegree; GWOdegree) – we used different out-degree terms in different models due to model convergence issues. A description of these variables is provided in Table 2.

**Table 2. Description of Network Variables**

| Variable | Description | Level of Variable |
|----------|-------------|-------------------|
| Edges | Density of the network | Social network process effect on tie formation |
| Mutual | Reciprocity of ties in the network | |
| DGWESP | Directed geometrically-weighted edgewise shared partner distribution. Transitivity in the network, i.e. "a friend of my friend also becomes a friend.". For directed networks triad type can be specified. | |
| GWIDegree | Geometrically-weighted in-degree distribution. Models the distribution of incoming ties in the network, i.e. popularity spread. | |
| GWODegree | Geometrically-weighted out-degree distribution. Models the activity spread of the network. | |
| ODegree | Distribution of outgoing ties in the network, i.e. activity spread. Fixed at specified values in the out-degree distribution. | |
| Edge.Cov | Multiplexity effect of a tie formed in a separate network on the probability of a tie forming in the current network. | Dyadic edge attribute effect on tie formation |
| Absdiff | Effect of the absolute difference between two nodes of a given node attribute (continuous variable), on the probability of a tie forming between a sender and receiver pair. Negative is homophily. Positive is heterophily. | Dyadic, paired-nodes attributes effect on tie formation |
| Diff | Effect of the difference between two nodes given a node attribute (conitunous) on the probability of tie formation. Useful for when the direction of difference is important. | |
| Nodematch | Effect of the similarity of two nodes of a given node attribute (categorical variable), on the probability of tie formation. Positive is homophily. Negative is heterophily. | |
| Receiver.Cov | Covariate effect of a node attribute (continuous variable) on the probability of *receiving* a tie. | Individual node attribute effect on tie formation |
| Sender.Cov | Covariate effect of a node attribute (continuous variable) on the probability of *sending* a tie. | |
| Receiver.Factor | Effect of a node attribute (categorical variable) on the probability of *receiving* a tie. | |
| Sender.Factor | Effect of a node attribute (categorical variable) on the probability of *sending* a tie. | |

## Method

To investigate the effects of the variables on information seeking behavior, we used exponential random graph modeling (ERGM). ERGM was developed to explore the factors that lead to the emergence of networks, permitting the inclusion of structural, dyadic, and actor-level attributes as predictors of tie formation in networks. In basic terms, the ERGM using Markov Chain Monte Carlo simulation to produce a probability distribution of simulated networks based on the observed network, providing estimates for model parameters. ERGM permits modeling the probability of tie formation, where the parameter estimates can be interpreted in a manner similar to logistic regression analysis. However, the advantage of ERGM is that it accounts for network dependency in probability of tie formation, which violates the assumption of independence of observations in the logistic regression setting. The ERGM model takes the following mathematical formulation (Robbins 2007).

$$\Pr(Y = y) = \left(\frac{1}{k}\right) exp \left\{ \sum_A \eta Ag \, A(y) \right\}$$

Actor attributes, dyadic homophily effects, and network structure effects are contained in gA(y). Structural effects control for dependencies in the network and can be thought of as standard network control variables that model properties of self-organization in the network. The model parameters are contained in ηA. The parameters undergo an iterative estimation and updating process until they effectively model the distribution of simulated networks. ERGM models converge when the observed network is probable given the simulated distribution of networks (Lusher, Koskinen and Robins 2013).

## Results

The results of the analysis begin with a brief examination of descriptive network statistics and visualization of the information search network within the city government. Figure 1 shows a directed network of 143 nodes (individuals) and 1778 edges (information search ties). The network is color coded by department and uses the Kamada-Kawai force directed layout to space the nodes and edges. The nodes are sized according to their degree. The visualization shows a robust and dense network of directed information search ties with strong departmental clustering.

**Figure 1 – Information Search Network**

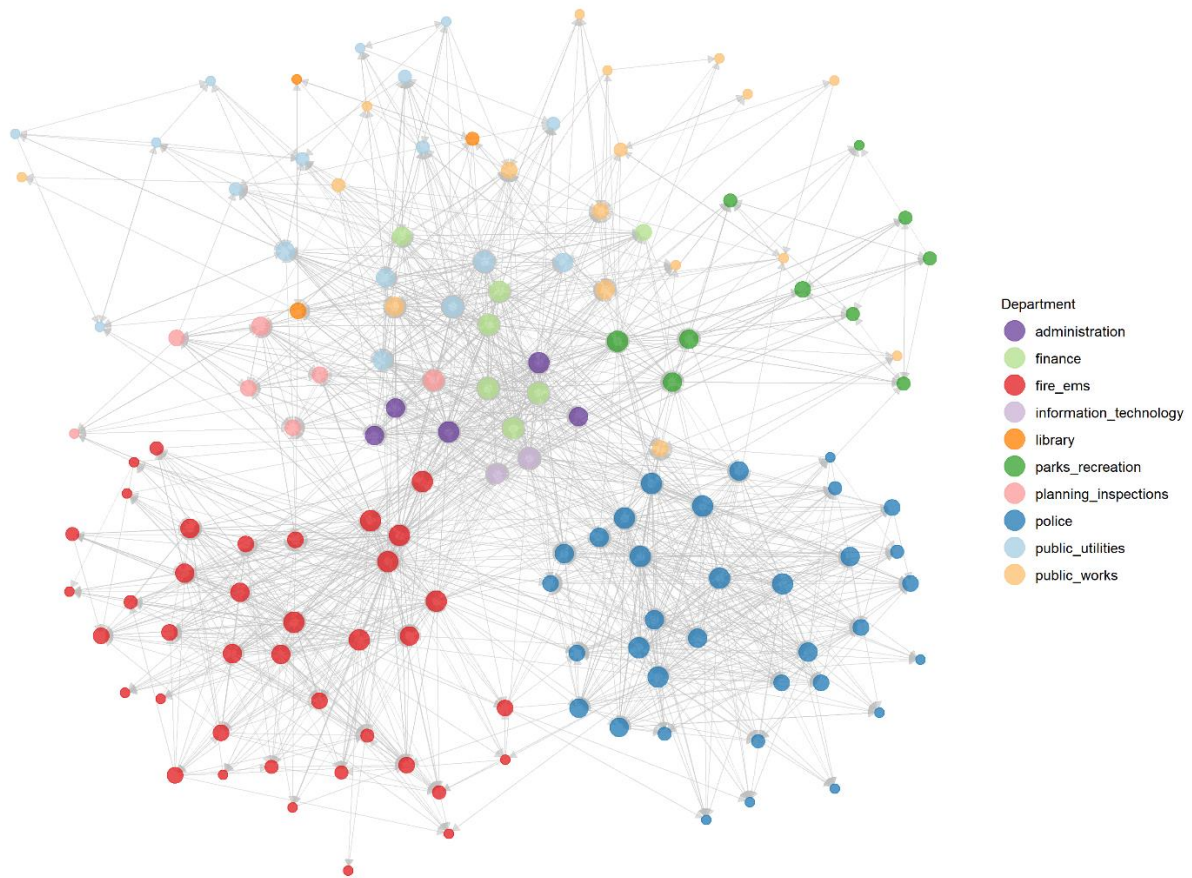

**Figure Notes:** The network visualization shows a directed network of individuals who listed others as individuals whom they seek for information to complete work-related tasks. The node color is based on the department. Node size is based on degree. Kamada-Kawai force directed algorithm was used for the network visualization layout. Two isolates were removed from the graph. Edges have arrows representing the directionality of the information search.

Table 3 shows descriptive statistics for the whole network and for individual node level values that are aggregated as the department mean. The whole network has a density of 0.09, a degree centralization score of 0.26, a transitivity score of 0.46, and reciprocity score of 0.26. Individuals within the Administration Department, Information and Technology Department, and the Finance Department have the highest centrality in the network relative to other departments. For example, Table 3 shows that comparing the average betweenness centrality scores of each department supports this observation: Administration has an average normalized betweenness centrality of 0.033, finance has a score of 0.019, and information and technology has a score of 0.016, while the average for the rest of the city is 0.007. The Police Department and the Fire/EMS Department display significant intra-departmental clustering relative to the rest of the network. For example, comparing the average clustering coefficient (transitivity score) of the departments supports this observation. Individuals within the fire/EMS department have an average transitivity score of 0.56, and police have a score 0.51, while the average for the rest of the departments is 0.346.

**Table 3. Descriptive Network Statistics for Full Network and Inter-Departmental Network**

| Whole Network Level | Nodes | Edges | Density | Centralization | Transitivity | Reciprocity |
|---|---|---|---|---|---|---|
|  | 143 / 143 | 1778 / 727 | 0.087 / 0.036 | 0.263 / 0.23 | 0.459 / 0.33 | 0.256 / 0.234 |

| Node Level Dept. Mean | Size | Intra-Dept. Ties | In-Degree | Out-Degree | B-Centrality | Transitivity |
|---|---|---|---|---|---|---|
| Administration | 5 | 0.184 | 20.6 / 16.8 | 28.8 / 25 | 0.033 / 0.032 | 0.262 / 0.199 |
| Finance | 7 | 0.216 | 19.86 / 15.6 | 24.29 / 20 | 0.019 / 0.015 | 0.252 / 0.164 |
| Fire & EMS | 39 | 0.86 | 11.23 / 1.6 | 12.87 / 3.23 | 0.010 / 0.002 | 0.557 / 0.102 |
| Info. & Technology | 2 | 0.026 | 39 / 38 | 10 / 9 | 0.016 / 0.028 | 0.301 / 0.202 |
| Library | 3 | 0.538 | 4 / 2 | 7.3 / 5.3 | 0.006 / 0.002 | 0.189 / 0.276 |
| Parks & Recreation | 10 | 0.491 | 11.5 / 5.9 | 9.7 / 4.1 | 0.011 / 0.004 | 0.386 / 0.138 |
| Planning & Inspections | 7 | 0.434 | 10.71 / 6.14 | 13 / 8.43 | 0.004 / 0.003 | 0.321 / 0.312 |
| Police | 34 | 0.806 | 14.41 / 2.82 | 14.15 / 2.56 | 0.008 / 0.001 | 0.513 / 0.168 |
| Public Utilities | 17 | 0.465 | 10.71 / 5.88 | 9.47 / 4.65 | 0.010 / 0.003 | 0.368 / 0.251 |
| Public Works | 19 | 0.281 | 6.58 / 4.84 | 3.63 / 1.9 | 0.004 / 0.002 | 0.370 / 0.255 |

Table Notes. The table shows descriptive statistics at the whole network level and at the node level aggregated by department mean. Statistics are included for the full network with all ties included (Figure 1), and for a separate network with all intra-departmental ties removed show next to the full value (Figure 3). Size is the same for both networks. Intra-Dept. Ties does not apply to the Inter-Departmental network.

Further supporting departmental clustering, a significant proportion of ties appear to be intra-departmental rather than between departments. Analyzing the departmental mixing matrix for the whole network showed that 59% of all 1778 information search ties are indeed within departments. However, this percentage is heavily skewed by the Police Department (81%, n=496) and the Fire/EMSDepartment (86%, n=442).

**Figure 2. Mixing Matrix Heatmap of Inter/Intra-Departmental Ties**

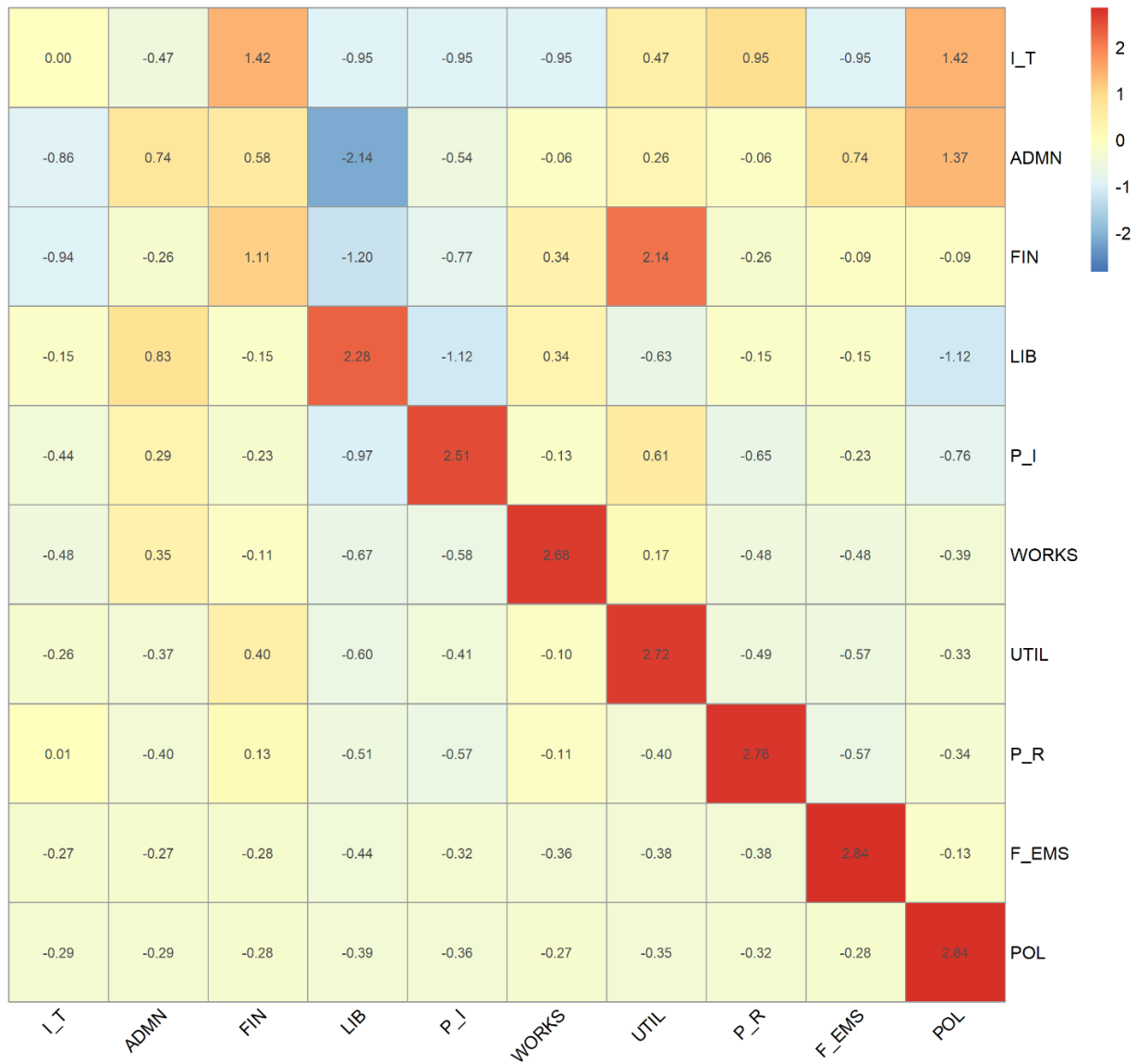

**Figure Notes.** The figure shows a heatmap of the departmental mixing matrix. Color is based on the normalized ratio intra and inter-departmental ties, where a darker red on the diagonal represents higher departmental homophily. Cells not on the diagonal show inter-departmental ties. The departments are ordered by strength of homophily along the diagonal. The matrix is directed, such that the Y axis represent ties from those departments to the departments on the X axis. Abbreviations -- I_I, information and technology; ADMIN, administration; FIN, finance; LIB, library; WORKS, public works; UTIL, public utilities; P_R, parks and recreation; F_EMS, fire and EMS; POL, police.

Figure 2 shows the mixing matrix of inter/intra-departmental ties, where the diagonal shows homophilous intra-departmental ties. The figure shows that Information and Technology, Administration, and Finance have a much lower ratio of homophilous ties, while also having more cross connections with various other departments, e.g. Finance sends a higher proportion of ties to Public Utilities than itself; Administration sends a higher proportion of ties to the Police Department. The opposite is true for the police and fire and EMS departments, where ties are mostly intra-departmental (see figure notes for more details).

**Figure 3. Inter-Departmental Information Search Network**

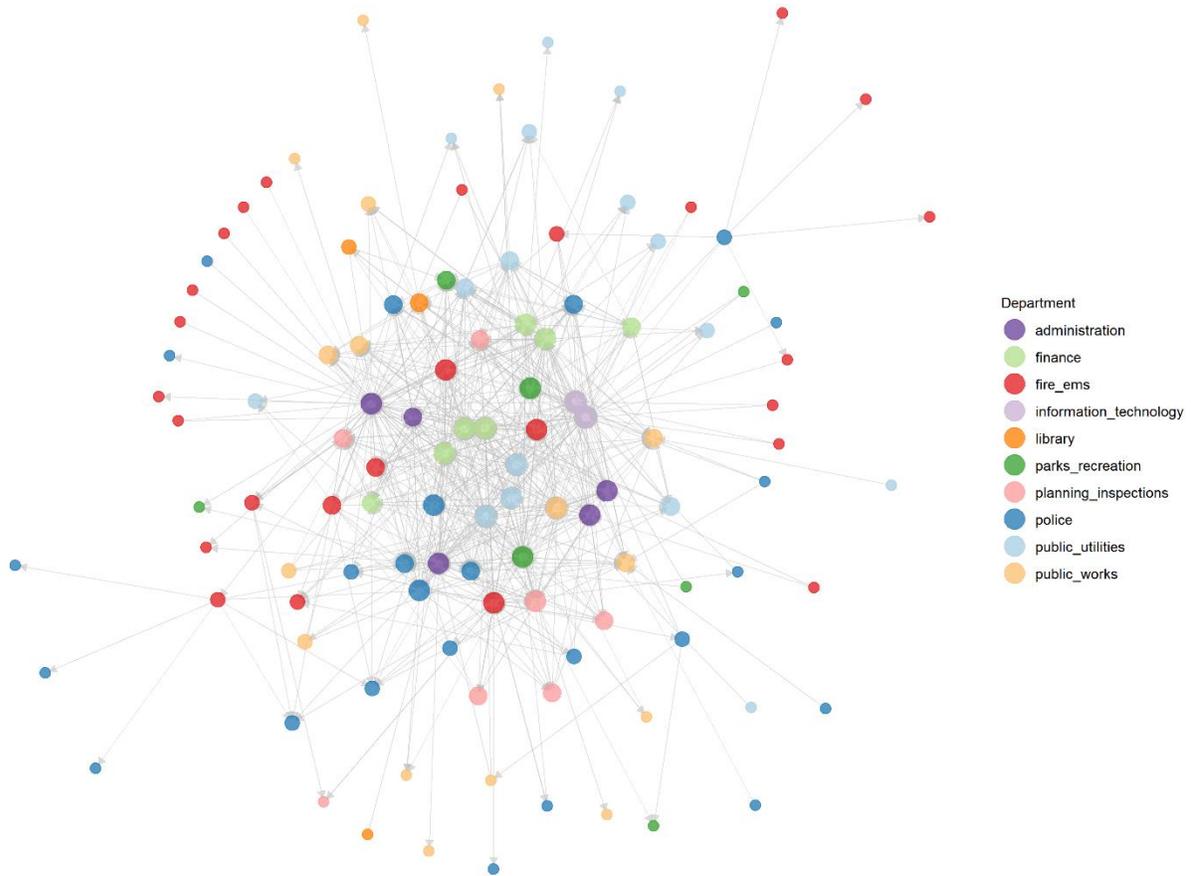

**Figure Notes.** The visualization shows a network of only cross-departmental ties, with all homophilous intra-departmental ties removed. Nodes are sized by degree and color coded by departmental membership. 29 isolated nodes were left out of the visualization.

Further, we observed that departments with taller hierarchies tended to have the highest proportion of homophilous ties. These departments were also much larger and had more ties overall. To focus on heterophilous departmental ties, we created a separate network that removed all intra-departmental ties. Figure 3 visualizes this network which has 727 edges instead of the full network which has 1778, with a density of 0.036, compared to 0.087 in the full network. The network clearly lacks departmental clustering, and it is far easier to see lateral ties across the whole organization. Table 3 shows much smaller indegree, outdegree, betweeness, and transitivity for homophilous departments in the inter-departmental network, relative to the full network.

Concluding the initial visual and descriptive analysis, variation in departmental centrality and clustering suggests that formal organizational structure affects the formation of the information search network, warranting further inferential analysis.

Moving beyond descriptive to inferential network analysis, Table 4 presents the results of the exponential random graph models (ERGM) performed on the full network (Model 1) and for the cross-departmental ties network (Model 2). Before examining the results, we discuss the model specification approach. The models include the effects of hierarchical status, supervisory status, permission seeking, and three attitudinal controls, including self-efficacy, trust, and commitment. Each variable includes a receiver effect, a sender effect, and an assortative mixing effect appropriate to the variable type (either nodematch, absdiff, or diff). We chose this approach in order to model the directionality of each variable effect. Categorical sender and receiver effects of department membership account for departmental size, hierarchy, etc. Nodematch for department (departmental homophily) was included only in Model 1, since Model 2 has no intra-departmental ties. Finally, both models include appropriate network controls that improve the goodness-of-fit and account for social network processes, including terms that model density, reciprocity, transitivity, popularity, and activity.

The model summary in Table 4 shows that individuals with higher hierarchical status (positioned further up the organizational hierarchy) tend to both receive more ties and send more ties in the information network, with a positive and significant estimate on both terms, providing support for H1. The Diff term for hierarchical status shows a negative and significant estimate, indicating that individuals lower in the hierarchy tend to seek out those that are higher.[7] Next, the model shows that individuals with supervisory status also tend to receive more ties in the network, providing further support for H1. However, the model did not produce a significant estimate for the effect of supervisors sending ties, or for a homophily effect between supervisors.

While, these effects appear to be accounted for by the more general hierarchical status variable,

it is interesting that both variables (status and supervisor) can achieve some degree of statistical

significance, even when tested together, pointing to the critical role of formal roles. Model 1

shows that the estimate on the permission edge covariate is positive and significant, indicating that the existence of a permission tie increases the probability of an information tie forming, providing support for H2. Further, the model shows that the estimate on departmental homophily is significant and positive, providing further support for H3. Indeed, departmental homophily appears to be the strongest predictor of an information search tie in Model 1.

    The estimates were mostly insignificant on the attitudinal controls included in Model 1,

with the exception of trust, which showed a positive and significant tie sender effect, self-efficacy, which showed a negative and significant receiver effect, and a homophily effect for commitment (a negative sign on abdiff is interpreted as homophily). Finally, five network control variables substantially improved the goodness-of-fit and support the presence of general social network processes, which must be accounted for in testing hypotheses regarding drivers of network ties within organizations.

---

[7] Additional models with alternative specifications are included in Online Appendix 4. Alternative models using the Absdiff term as opposed to the Diff term show that there is a more general tendency for ties to form between individuals of similar hierarchical status; there is a significant interaction between hierarchical status and supervisory status; and both departmental size and hierarchical levels within the department had negative effects on tie formation.

**Table 4 – Exponential Random Graph Models**

| Model Terms | Model 1 (All Ties) | Model 2 (Inter-Dept. Ties) |
|---|---|---|
| Hierarchical.Status - Receiver.Cov | 0.331 (0.049)*** | 0.372 (0.076)*** |
| Hierarchical.Status - Sender.Cov | 0.300 (0.043)*** | 0.266 (0.066)*** |
| Hierarchical.Status - Diff | -0.480 (0.081)*** | -0.174 (0.118) |
| Supervisor - Receiver.Factor | 0.459 (0.079)*** | 0.215 (0.095)* |
| Supervisor - Sender.Factor | 0.035 (0.056) | -0.087 (0.087) |
| Supervisor - Nodematch | 0.067 (0.065) | 0.185 (0.078)* |
| Permission - Edge.Cov | 0.959 (0.126)*** | 0.723 (0.166)*** |
| Department - Sender.Factor | FIXED | FIXED |
| Department - Receiver.Factor | FIXED | FIXED |
| Department - Nodematch | 3.505 (0.112)*** | |
| Self.Efficacy - Receiver.Cov | -0.166 (0.044)*** | -0.110 (0.055)* |
| Self.Efficacy - Sender.Cov | -0.059 (0.035) | -0.138 (0.052)** |
| Self.Efficacy - Absdiff | -0.064 (0.039) | -0.093 (0.051) |
| Trust - Receiver.Cov | -0.005 (0.046) | 0.017 (0.057) |
| Trust - Sender.Cov | 0.097 (0.031)** | 0.106 (0.053)* |
| Trust - Absdiff | -0.040 (0.039) | 0.037 (0.051) |
| Commitment - Receiver.Cov | -0.023 (0.042) | 0.024 (0.052) |
| Commitment - Sender.Cov | -0.007 (0.030) | 0.071 (0.046) |
| Commitment - Absdiff | -0.101 (0.043)* | -0.023 (0.054) |
| Density - Edges | -4.556 (0.305)*** | -5.963 (0.411)*** |
| Reciprocity - Mutual | 0.742 (0.116)*** | 1.880 (0.162)*** |
| Transitivity - DGWESP.OSP.0.75 | 0.182 (0.042)*** | |
| Popularity - GWIDegree.0.75 | -1.036 (0.294)*** | |
| Activity - ODEGREE.0:5 | FIXED | |
| Transitivity - DGWESP.OTP.0.75 | | 0.289 (0.065)*** |
| GWIDegree.0.25 | | -1.971 (0.313)*** |
| GWODegree.025 | | -2.928 (0.326)*** |
| AIC | 7281.972 | 3893.721 |
| BIC | 7638.312 | 4202.549 |
| Log Likelihood | -3595.986 | -1907.86 |

**Table Notes**: ***p < 0.001; **p < 0.01; *p < 0.05. Standard errors are in parentheses. Dependent variable is the probability of a tie forming between two nodes in the network. Diff options included, pow=0, which avoids overspecification, reducing the values to 1 for a positive difference, 0 for no difference, and -1 for a negative difference between the i and j nodes. DGWESP options include 0.75 fixed alpha for both models, and triad type OSP for Model 1 and OTP for Model 2. GWIDegree includes a fixed alpha at 0.75 for Model 1 and 0.25 for Model 2. ODegree is fixed at 0 through 5 out-degrees, and each term is significant (abbreviated as FIXED). GWODegree for Model 2 uses a fixed alpha of 0.25. For an explanation of each model term see Table 2. ERGM controls for both models include seed=101; MCMC.samplesize=5000; MCMC.interval=5000; MCMC.burnin = 80000. Goodness-of-fit plots for the model are shown in Online Appendix 3. Exact P-Values are also shown in Online Appendix 3. R code is available upon request.

Next, Model 2, which contains only the subset of inter-departmental ties, shows a similar pattern of drivers of information search to Model 1. A comparison of models shows some statistical differences. First, the tendency for lower hierarchical status individuals to reach upward is rendered insignificant when focusing only on cross-departmental ties. Second, homophily between supervisors becomes significant indicating that supervisors may be more likely to form lateral ties with each other.[8]

In summary, the results of the ERGM analysis suggest that formal organizational structure has important effects on the formation of information search networks within city government, even while controlling for well-recognized perception-based factors as well as social network processes. The results suggest broad support for all three hypotheses.

## Discussion

This article sought to develop and test theory regarding the effects of formal organizational structure on the development of informal social networks within public organizations. The results showed broad support for the subject-line expectation, providing empirical evidence for the importance of a number of measures of formal organizational structure. To summarize, we found that formal status has important effects on information search, where those with higher hierarchical status tend to be both more active and more popular targets for information search, while individuals lower in the hierarchy tend to search upward.

[8] The network statistics on Model 2 displayed a similar pattern to Model 1. However, we were able to use less idiosyncratic modeling terms in Model 2, due to eliminating a large number of intra-departmental triads. Thus, the model fit for Model 2 is better than Model 1 (see Appendix 3 for GOF plots).

We also found evidence for multiplexity in the network, where information search ties tend to co-occur with permission search ties. Finally, we found that information search is heavily shaped by both department specific variability and by a more general pattern of departmental homophily.

Our findings regarding the effects of hierarchical power on network activity speaks to a body of research on informal networks (Krackhart, 1990; Krackhardt and Hanson, 1993; Kilduff and Krackhardt's 2008) that analyzed the influence of formal status on informal network development. Those located higher in the organization are better positioned to observe and influence informal networks. While we found that supervisors appeared to be the recipients of more ties, those with close proximity to power, controlling for supervisor status, were more active in seeking information both within and between departments (supporting Kahn and Kram 1994). Further, the fact that managers high up in the hierarchy are likely to be sought out for information emphasizes a dilemma: While supportive managers may want to play this role and provide insights and feedback to employees, organizations need to establish structures that prevent managers from being overburdened or creating a bottleneck. As Maroulis, Diermeier, and Nisar (2020, 77) suggest, managers can enhance informational diversity within organizational sub-units by reconfiguring them to contain more cross-unit ties. One way to do this might be to establish staff positions, which could serve as a filter for all coordination and communication related inquiries. A second possibility would be the creation of learning forums around themes such as improving performance or visualizing impact (Moynihan 2005), which would allow managers to interact with a cross-section of employees, while being focused on specific issues for a limited time period.

Our finding for the permission network has two implications. First, the lines of permission matter a great deal even for the social and largely informal phenomenon of information seeking. Second, while permission pathways are one important structural factor, it is not the only one. Put another way, even after controlling for the permission network, all other structural-formal factors, including hierarchical status, supervisor status, and departmental membership, still show significant, independent effects. This finding confirms our point that the structural side of organizations has several facets, and many of them come into play at the same time.

Scholars have argued that hierarchy should have a waning influence over information seeking, presumably due to organizations becoming flatter and less layered, thus creating the need for organizational members to actively seek information outside chains of command. In support of this argument, Cross, Rice and Parker (2001) found that being at the same hierarchical level did not influence information benefits, nor did colocation in the same office. Yet, our findings suggest the hierarchy significantly shapes information flows within organizations, constraining them both vertically and horizontally. Our finding of strong homophily based on departmental structure supports Siciliano's (2015) previous finding that schoolteachers display grade level homophily. Interestingly, Nisar and Maroulis (2017) find no such homophily effect in the presence of a major organizational reform initiative. This suggests that when homophily becomes undesirable, the structure itself may be altered to stimulate the development of information search pathways across organizational sub-structures. Thus, public organizations wishing to foster information flow and stimulate collaboration across departments should consider intentional cross-departmental initiatives to overcome the impeding influence of formal structure on organization-wide information flow. Given the role of departments in constraining information flow, public managers should consider explicitly encouraging cross-departmental information exchange that will diminish silos.

While a great deal of research on information search and knowledge sharing has been concerned with the bridging of organizational silos and facilitating lateral communication, our findings point to one additional consideration: If department membership is still one of the main factors for explaining information exchange, then one way to further improve communication is through well-developed relations *within* work units. This finding is in line with research that has argued that the development of social capital within teams can benefit the organization as a whole (Kroll et al. 2019). Information sharing within teams or departments could be fostered if members are encouraged to speak up and listen to each other and feel safe to take risks (Edmondson and Roloff 2008).

Further, we found that accounting for formal structure had important consequences for existing perceptual-cognitive variables. The results suggested that trust is strongly associated with information search regardless of the model specification or network configuration, while organizational commitment appears to be more relevant at the cross-departmental level and loses much of its significance after controlling for variables that capture organizational structure and social network processes. Interestingly, self-efficacy, which had displayed mixed significance and direction of effect on network ties in previous studies, shows a strong significant negative effect on information search in our study. This finding fits with Binz-Scharf, Lazer, and Mergel's (2012) observation that reputational concerns remain a "major obstacle" to information sharing in public organizations.

Like other research, our study is prone to some limitations. First, the analysis is cross-sectional and does not account for change over time. Rather, the study utilizes ERGM to analyze a snapshot of a social network to test hypotheses regarding the formation of social ties. Future research could extend the study to account for temporal dynamics. Second, the study includes only one organization. Hence, we cannot generalize the findings far beyond the present context. However, we take a different philosophical approach, emphasizing the current study as an instrumental case for the development of theory and its illustration, rather than generalizing to a broader population. Analysis of public sector organizations with different department structures, of different sizes, in different geographic locations, may produce valuable additional insights. Future research could extend the study to account for more organizations. Recently developed, multilevel ERGMS could be used in future research to control for departmental clustering in intraorganizational networks (see Stewart et al. 2019).

## Conclusion

The vast majority of research on public sector networks has focused on the antecedents and consequences of *inter*-organizational networks, often neglecting *intra*-organizational social networks. Those studies that have examined intra-organizational networks have tended to focus on the effect of informal dynamics of social structure, cognitive-perceptual variables, and the attributes of individuals, such as commitment, trust, and self-efficacy. As a result, the function of formal organizational structure in determining the shape of informal intra-organizational networks has been relatively neglected. The current study addresses this lacuna by developing theory and testing hypotheses regarding the effects of formal structure on information search within a small city government. The results suggest strong effects of formal structure on constraining and enabling information search across and within departments, as well as interesting relative effects of trust, commitment, and self-efficacy in the presence and absence of formal structural variables.

Our call to better integrate formal organization and informal social structures in public administration research mirrors similar developments and emerging research in other disciplines (Hunter et al. 2020; McEvily et al. 2014; Spillane and Kim 2012). To that end, we show that the literature on formal organization bears a great deal of potential to further enhance theories of social structure, using information search as our phenomenon of interest. Empirically, we find support for the important role of formal status, permission pathways, and departmental membership. We draw attention to generalist, structurally fragmented public organizations (such as local governments) for which formal organization tends to be particularly instructive in shaping organizational behavior. Our call for more theoretical integration goes beyond the use of formal-structural factors as controls. Rather, we see much potential in theorizing, where i) formal structure is used to identify contingency factors for social network theory; ii) formal and informal structures reinforce or mitigate each other's impact on behavioral outcomes; or iii) both types of structure directly influence each other, either in a complementary or substitutional manner (see also McEvily et al. 2014).

Considering that bureaucracies are known for their reliance on tall hierarchies, formal authority, and routine processes, we suggest that these factors will also be influential in shaping more informal, social interactions. This is not to say that formal structure would be the only factor that requires more attention. Organizational culture, for example, remains a salient feature of tall organizations, as well as tall organizational sub-units, e.g., police departments, which can further shape the development of informal networks in unique ways. The incorporation of organizational culture into social network analysis is thus a potential avenue for future research.

In addition, our analysis showed strong differences between departments within the same organization, pointing to the importance of task variables and, possibly, culture differences that may even exist at the sub-organizational level. If one's research interest is in understanding behaviors within public organizations, then the mere use of variables that are widely employed in the analysis of firms (such as network position, work attitudes, or demographic attributes) may be insufficient. Overall, our study provides a practical starting point for the unification of formal and informal network structure within public sector intra-organizational network studies.

**Acknowledgements:** We would like to thank Michael Siciliano and three anonymous reviewers for their useful comments on the manuscript.

**Data Availability Statement:** The data underlying this article are available on figshare, *at* the following DOI URLs.

https://doi.org/10.6084/m9.figshare.13714063.v1
https://doi.org/10.6084/m9.figshare.13714060.v1
https://doi.org/10.6084/m9.figshare.13713895

# Online Appendices

## Online Appendix 1 – Survey Items

The table below shows the exact language used for each survey item. Each variable was measured on a 1-7 Likert scale.

Table A.1 – Interview Items

| Variable | Questionnaire Text | Operationalization |
|---|---|---|
| Information Search Tie | "I seek information from this person to do my job" | Dependent Variable- Serves as edge between two nodes in information search network |
| Self-Efficacy 1 | "I am confident that I can perform effectively on many different tasks." Likert 1-7 | Independent Variable – Self-Efficacy Factor |
| Self-Efficacy 2 | "Even when things are tough, I can perform quite well." Likert 1-7 | |
| Self-Efficacy 3 | "Compared to other people, I can do most tasks very well." Likert 1-7 | |
| Trust 1 | "In my department, employees trust supervisors." Likert 1-7 | Independent Variable – Trust Factor |
| Trust 2 | "In my department, supervisors trust their subordinates." Likert 1-7 | |
| Trust 3 | "In my department, employees trust supervisors to make good decisions." Likert 1-7 | |
| Commitment 1 | "I would feel guilty if I left this organization now." Likert 1-7 | Independent Variable – Commitment Factor |
| Commitment 2 | "I would be very happy to spend the rest of my career with this organization." Likert 1-7 | |
| Commitment 3 | "I am thinking about leaving this organization." Likert 1-7 | |
| Permission search tie | "I seek permission from this person to do certain tasks" Likert 1-7 | Independent Variable –Serves as edge between two nodes in permission search network |

**Online Appendix 2. Factor Analysis**

Table A.2 – Factor Loadings

| Items | Factor 1 | Factor 2 | Factor 3 |
|---|---|---|---|
| Self-Efficacy 1 | | 0.93 | |
| Self-Efficacy 2 | | 0.95 | |
| Self-Efficacy 3 | | 0.92 | |
| Trust 1 | 0.92 | | |
| Trust 2 | 0.97 | | |
| Trust 3 | 0.88 | | |
| Commitment 1 | | | 0.83 |
| Commitment 2 | | | 0.87 |
| Commitment 3 | | | 0.92 |
| Eigenvalue | 4.18 | 2.05 | 1.34 |

Table Notes. Shows factor loadings based on principal component factoring. Loadings < 0.4 were omitted.

**Online Appendix 3 – Goodness of Fit Statistics for ERGMs**

The package *statnet* in R produces the goodness-of-fit plots, presented in Figure A.3.1 and A.3.2 for Table 4 Model 1 and Model 2. The models converged to produce reliable parameter estimates. The figures show how well the distribution of simulated networks models observed distributions of interest, including the in-degree distribution, out-degree distribution, edgewise shared partner distribution, network distance, and model parameters. The figures show a reasonably good fit given the unevenness of the in-degree and out-degree distributions in Model 1. The ODegree term was useful for fixing the distribution between 0 and 5 degrees where drastic changes were observed in the distribution. We were unable to increase the DGWESP fixed alpha value or include another triangle term such as GWDSP without model degeneracy in Model 1.

**Figure A.3.1 – ERGM Goodness-of-Fit Diagnostic Plots, Model 1, Table 4**

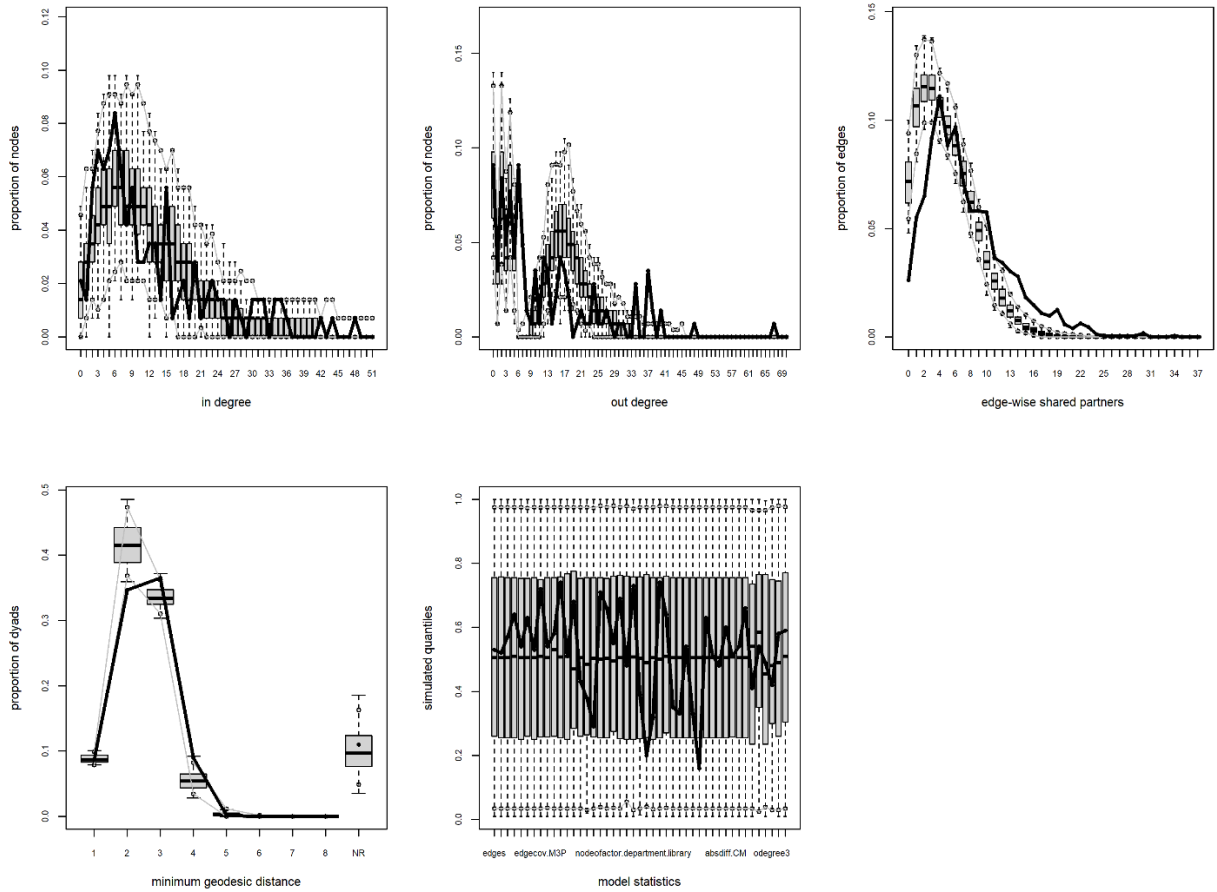

Figure Notes: The plots include an in-degree distribution, an out-degree distribution, an edgewise shared partner distribution, a distance distribution, and a set of model parameter box plots. The solid line in each figure represents the observed distributions, while the box plots represent the simulated distributions based on observed network characteristics and model terms. The degree to which the box plots encompass the black line indicates the goodness-of-fit for the model for the given parameter.

**Figure A.3.2 - ERGM Goodness-of-Fit Diagnostic Plots, Model 2, Table 4**

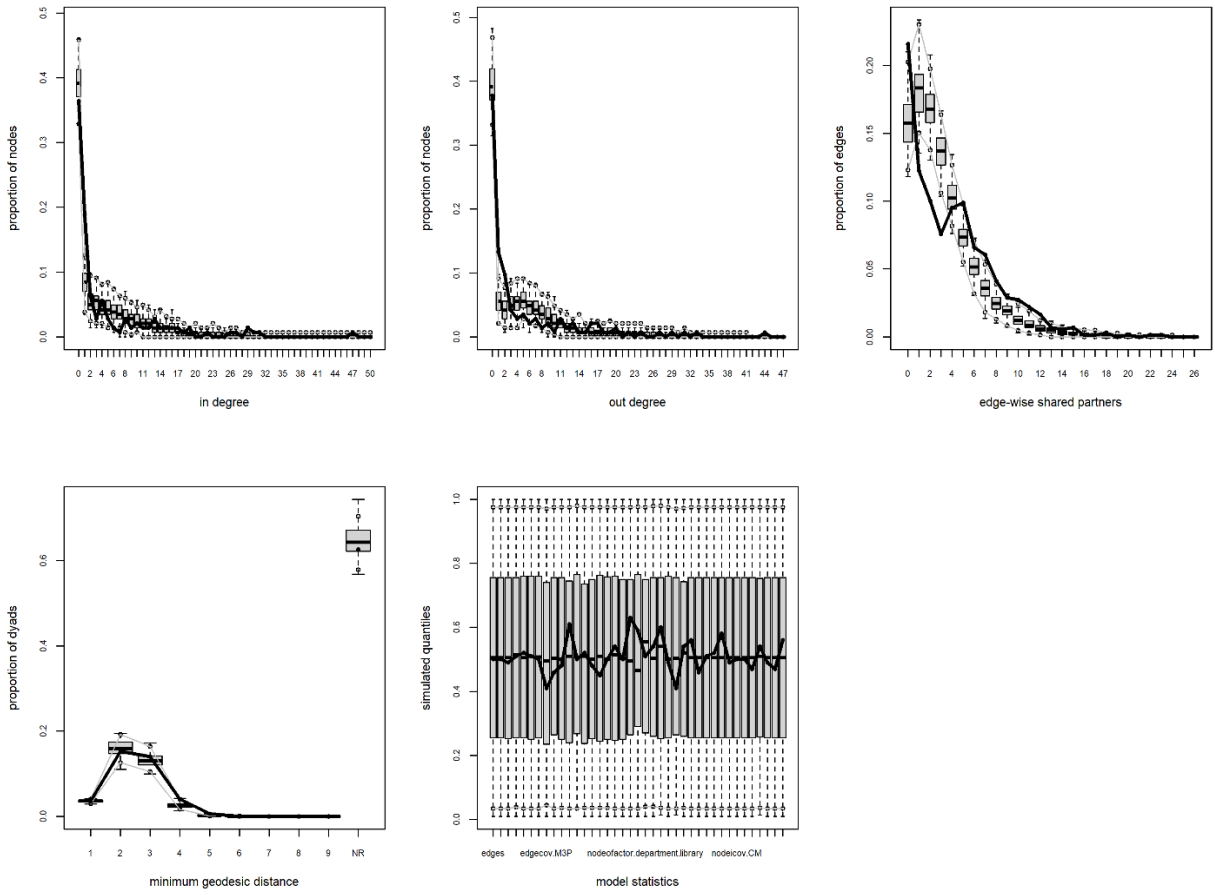

**Table A.3.1 – Exact P-Values Corresponding to Estimates and Standard Errors in Table 4**

| Model Terms | Model 1 (All Ties) | Model 2 (Inter-Dept. Ties) |
|---|---|---|
| Hierarchical.Status - Receiver.Cov | 2.114359e-11 | 9.161165e-07 |
| Hierarchical.Status - Sender.Cov | 4.042116e-12 | 5.346552e-05 |
| Hierarchical.Status - Diff | 3.107222e-09 | 1.401903e-01 |
| Supervisor - Receiver.Factor | 7.298388e-09 | 2.318757e-02 |
| Supervisor - Sender.Factor | 5.322027e-01 | 3.185137e-01 |
| Supervisor - Nodematch | 2.988411e-01 | 1.742608e-02 |
| Permission - Edge.Cov | 2.835701e-14 | 1.291068e-05 |
| Department - Sender.Factor | FIXED | FIXED |
| Department - Receiver.Factor | FIXED | FIXED |
| Department - Nodematch | 2.780862e-213 | |
| Self.Efficacy - Receiver.Cov | 1.522995e-04 | 4.373481e-02 |
| Self.Efficacy - Sender.Cov | 8.841529e-02 | 7.294600e-03 |
| Self.Efficacy - Absdiff | 1.033555e-01 | 6.543857e-02 |
| Trust - Receiver.Cov | 9.091278e-01 | 7.627468e-01 |
| Trust - Sender.Cov | 1.966306e-03 | 4.485997e-02 |
| Trust - Absdiff | 3.052078e-01 | 4.677894e-01 |
| Commitment - Receiver.Cov | 5.804056e-01 | 6.493477e-01 |
| Commitment - Sender.Cov | 8.022395e-01 | 1.217654e-01 |
| Commitment - Absdiff | 1.863455e-02 | 6.778840e-01 |
| Density - Edges | 1.492697e-50 | 9.061495e-48 |
| Reciprocity - Mutual | 1.355996e-10 | 4.260361e-31 |
| Transitivity - DGWESP.OSP.0.75 | 1.669317e-05 | |
| Popularity - GWIDegree.0.75 | 4.167414e-04 | |
| Activity - ODEGREE.0:5 | FIXED | |
| Transitivity - DGWESP.OTP.0.75 | | 8.690350e-06 |
| GWIDegree.0.25 | | 3.177262e-10 |
| GWODegree.025 | | 2.425442e-19 |

## Online Appendix 4 – Alternative Model Specifications

Table A.4 includes alternative models specifications based on the full network that may be of interest to readers. Model 1 includes interaction terms between supervisor and hierarchical status. Model 2 uses Absdiff for hierarchical status rather than Diff. Model 3 drops categorical node effect for department and includes department size. Model 4 includes the absolute level of hierarchy for the department.

Table A.4 – Alternative Models

| | Model 1 | Model 2 | Model 3 | Model 4 |
|---|---|---|---|---|
| edges | -3.323 (0.356)*** | -3.744 (0.323)*** | -4.288 (0.154)*** | -5.270 (0.253)*** |
| nodeicov.hierarchical_status | 0.175 (0.048)*** | 0.403 (0.034)*** | 0.144 (0.041)*** | 0.269 (0.047)*** |
| nodeocov. hierarchical_status | 0.347 (0.042)*** | 0.156 (0.027)*** | 0.242 (0.037)*** | 0.310 (0.042)*** |
| diff0.t-h. hierarchical_status | -0.516 (0.077)*** | | -0.408 (0.073)*** | -0.405 (0.077)*** |
| nodeifactor.supervisor.1 | 0.147 (0.198) | 0.646 (0.071)*** | 0.633 (0.063)*** | 0.529 (0.068)*** |
| nodeofactor.supervisor.1 | -0.570 (0.140)*** | -0.049 (0.054) | 0.017 (0.052) | 0.005 (0.054) |
| nodematch.supervisor | -0.450 (0.152)** | 0.036 (0.061) | 0.015 (0.056) | 0.005 (0.058) |
| nodeicov.hierarch*super | 0.147 (0.051)** | | | |
| nodeocov. hierarch*super | 0.164 (0.039)*** | | | |
| absdiff.hierarch*super | -0.145 (0.038)*** | | | |
| edgecov.M3P | 0.934 (0.122)*** | 0.957 (0.121)*** | 0.962 (0.111)*** | 0.904 (0.119)*** |
| nodematch.department | 3.480 (0.108)*** | 3.337 (0.109)*** | 2.927 (0.082)*** | 2.680 (0.087)*** |
| nodefactor.department | fixed | fixed | | |
| nodeicov.SE | -0.155 (0.043)*** | -0.177 (0.042)*** | -0.173 (0.039)*** | -0.181 (0.042)*** |
| nodeocov.SE | -0.048 (0.031) | -0.077 (0.034)* | -0.092 (0.031)** | -0.099 (0.035)** |
| absdiff.SE | -0.055 (0.038) | -0.067 (0.038) | -0.112 (0.035)** | -0.096 (0.037)** |
| nodeicov.TR | -0.054 (0.044) | -0.038 (0.044) | -0.055 (0.039) | -0.067 (0.043) |
| nodeocov.TR | 0.107 (0.030)*** | 0.153 (0.033)*** | 0.123 (0.030)*** | 0.149 (0.034)*** |
| absdiff.TR | -0.046 (0.039) | -0.039 (0.041) | 0.086 (0.034)* | 0.097 (0.036)** |
| nodeicov.CM | -0.037 (0.038) | -0.037 (0.038) | -0.064 (0.033) | -0.067 (0.035) |
| nodeocov.CM | 0.009 (0.028) | 0.016 (0.030) | -0.010 (0.027) | 0.006 (0.028) |
| absdiff.CM | -0.120 (0.041)** | -0.115 (0.042)** | -0.231 (0.034)*** | -0.241 (0.038)*** |
| mutual | 0.643 (0.108)*** | 0.587 (0.107)*** | 0.741 (0.100)*** | 0.858 (0.104)*** |
| gwesp.OSP.fixed.0.75 | 0.173 (0.044)*** | 0.211 (0.046)*** | 0.341 (0.043)*** | 0.284 (0.046)*** |
| gwideg.fixed.0.75 | -1.328 (0.353)*** | -1.002 (0.370)** | 0.030 (0.448) | 0.298 (0.431) |
| odegree(0:5) | fixed | fixed | fixed | fixed |
| absdiff.hierarchical_status | | -0.151 (0.027)*** | | |
| nodeicov.deptsize | | | -0.017 (0.003)*** | |
| nodeocov.deptsize | | | -0.020 (0.002)*** | |
| nodeicov.dept.hierarchy | | | | -0.009 (0.031) |
| nodeocov.dept.hierarchy | | | | -0.100 (0.028)*** |
| AIC | 7359.332 | 7387.3 | 7650.196 | 7780.545 |
| BIC | 7668.16 | 7672.373 | 7879.838 | 8010.187 |
| Log Likelihood | -3640.666 | -3657.65 | -3796.098 | -3861.273 |

***p < 0.001; **p < 0.01; *p < 0.05